\documentclass{jps-cp}
\usepackage{txfonts} 
\usepackage{bm}

\title{Three-loop corrections to the quark and gluon decomposition of the QCD trace anomaly and their applications}

\author{Kazuhiro \textsc{Tanaka}}

\inst{Department of Physics, Juntendo University, Inzai,
  Chiba 270-1695, Japan
}

\email{kztanaka@juntendo.ac.jp}

\recdate{March 25, 2022}

\abst{In the QCD energy-momentum tensor $T^{\mu\nu}$, the terms that contribute to physical matrix elements are expressed as the sum of the gauge-invariant quark part and gluon part. Each part undergoes the renormalization due to the interactions among quarks and gluons, although the total tensor $T^{\mu\nu}$ is not renormalized thanks to conservation of energy and momentum. We show that, through the renormalization, each of the quark and gluon parts of $T^{\mu\nu}$ receives a definite amount of anomalous trace contribution, such that their sum reproduces the well-known QCD trace anomaly. We provide a procedure to derive such anomalous trace contribution for each quark/gluon part to all orders in perturbation theory, and obtain the corresponding explicit formulas up to three-loop order in the 
$\overline{\rm MS}$
scheme in the dimensional regularization. We apply our three-loop formulas of the quark/gluon decomposition of the trace anomaly to calculate the anomaly-induced mass structure of nucleons as well as pions. Another application of our three-loop formulas is a quantitative analysis for the constraints on the twist-four gravitational form factors 
of the nucleon, $\bar{C}_{q,g}$.
}

\kword{trace anomaly, energy-momentum tensor, QCD, gravitational form factors}

\begin{document}
\maketitle

\section{Introduction}
The QCD energy-momentum tensor $T^{\mu\nu}$ 
is known to receive the trace anomaly~\cite{Collins:1976yq} as 
\begin{equation}
T^\mu_\mu=\eta_{\mu\nu} T^{\mu\nu} =\frac{\beta(g)}{2g}F^2
+ \left(1+\gamma_m(g)\right)m\bar{\psi}\psi\ ,  
\label{111}
\end{equation}
representing the broken scale invariance due to the quantum loop effects,
with the beta-function $\beta$ for the QCD coupling constant $g$ 
and the anomalous dimension $\gamma_m$ for the quark mass $m$. 
Here, $\eta_{\mu\nu}$ is the 
the metric tensor, and 
$F^2$ ($=F_a^{\mu\nu}{F_a}_{\mu\nu}$) and $\bar{\psi}\psi$ denote the renormalized composite operators dependent on a renormalization scale.
The symmetric energy-momentum tensor $T^{\mu\nu}$ is expressed as the sum 
of the gauge-invariant quark part $T^{\mu\nu}_q$ and gluon part $T^{\mu\nu}_g$, as ($D^\mu=\partial^\mu+ig A^\mu$, $R^{(\mu}S^{\nu)}\equiv \left(R^\mu S^\nu+R^\nu S^\mu\right)/2$)
\begin{equation}
T^{\mu\nu} = \frac{1}{2}\bar{\psi} \gamma^{(\mu} i\overleftrightarrow{D}^{\nu )} \psi+ \left(F_a^{\mu\rho}{{F_a}_\rho}^\nu+\frac{\eta^{\mu \nu}}{4}F_a^{\lambda\rho}{F_a}_{\lambda\rho}\right)\equiv   T^{\mu\nu}_q+ T^{\mu\nu}_g \ ,
\label{tqg}
\end{equation}
using the QCD equations of motion (EOM),
up to the ghost and gauge-fixing terms that are not relevant for the following discussion.
Classically, we have, 
${\eta_{\mu \nu }}T_q^{\mu \nu } = m\bar \psi\psi$ and ${\eta_{\mu \nu }}T_g^{\mu \nu } = 0$, up to the terms that vanish by the EOM, 
but (\ref{111}) does not coincide with the quantum corrections to the $m\bar \psi\psi$ operator,
reflecting that renormalizing the quantum loops and taking the trace do not commute.
We note that the total tensor $T^{\mu\nu}$ of (\ref{tqg}) is not renormalized; it is a finite, scale-independent operator, because of
the energy-momentum conservation, 
$\partial_\nu T^{\mu\nu}=0$,
while $T_q^{\mu\nu}$ and $T_g^{\mu\nu}$ are not conserved separately 
and $T_q^{\mu\nu}$ as well as $T_g^{\mu\nu}$ is subject to regularization and renormalization.
This fact suggests that each of $T_q^{\mu\nu}$ and $T_g^{\mu\nu}$ should
receive
a definite amount of anomalous trace contribution, such that their sum reproduces~(\ref{111}).
The corresponding trace anomaly for each quark/gluon part
is derived up to two-loop order in \cite{Hatta:2018sqd}. The extension to the three-loop order
is worked out in \cite{Tanaka:2018nae}, 
demonstrating that the logic to determine the quark/gluon decomposition of the trace anomaly holds to all orders in perturbation theory.
In the MS-like (MS,  $\overline{\rm MS}$) schemes in the dimensional regularization,  we obtain
\begin{equation}
\eta_{\mu \nu }T_q^{\mu \nu }  =  m\bar \psi\psi + \frac{{{\alpha _s}}}{{4\pi }}\left( {\frac{{{n_f}}}{3}{F^2} + \frac{{4{C_F}}}{3}m\bar \psi\psi} 
\right)+\cdots\ , \;\;\;\;\;
\eta_{\mu \nu }T_g^{\mu \nu }  = \frac{{{\alpha _s}}}{{4\pi }}\left( { - \frac{{11{C_A}}}{6}{F^2} + \frac{{14{C_F}}}{3}m\bar \psi\psi} \right)+\cdots\ ,
\label{ano}
\end{equation}
for $n_f$ flavor and $N_c$ color with $C_F=(N_c^2-1)/(2N_c)$ and $C_A=N_c$, 
where the ellipses stand for the two-loop (${\cal O}(\alpha_s^2)$)
as well as three-loop  (${\cal O}(\alpha_s^3)$) corrections, whose explicit formulas are presented in \cite{Hatta:2018sqd,Tanaka:2018nae}.
The sum of the two formulas of (\ref{ano}) coincides with (\ref{111})
at every order in $\alpha_s$ ($=g^2/(4\pi)$).

\section{Renormalization mixing at three loop}  

We sketch how the formulas (\ref{ano}) 
are obtained. 
First of all, the renormalization of $T_q^{\mu\nu}$, $T_g^{\mu\nu}$
of (\ref{tqg}) 
is not straightforward. Indeed, $T_q^{\mu\nu}$, $T_g^{\mu\nu}$ are composed of 
the twist-two (traceless part) and twist-four (trace part) operators
and
the renormalization mixing between the quark part and gluon part also arises.
To treat them, we define a basis of independent gauge-invariant operators up to twist four,
\begin{eqnarray}
O_q=i\bar{\psi}\gamma^{(\mu}\overleftrightarrow{D}^{\nu)} \psi\ ,  \;\;\;\;\;\;\;\;\;\;\;\;
O_{q(4)}=\eta^{\mu\nu}m\bar{\psi}\psi\ ,  \;\;\;\;\;\;\;\;\;\;\;\;
O_g=-F^{\mu\lambda}F^{\nu}_{\ \lambda}\ ,  \;\;\;\;\;\;\;\;\;\;\;\;
O_{g(4)}= \eta^{\mu\nu}F^2\ ,
\end{eqnarray}
and the corresponding bare operators, $O_k^B$.
The renormalization constants are introduced as
\begin{eqnarray}
&&O_g=Z_T O_g^B + Z_MO_{g(4)}^B + Z_L O_q^B + Z_S O_{q(4)}^B\ , \;\;\;\;\;\;
O_q=Z_\psi O_q^B +Z_KO_{q(4)}^B+ Z_Q O_g^B +Z_B O_{g(4)}^B\ ,  \label{o3ren}\\
&& O_{g(4)}=Z_F O_{g(4)}^B+Z_C O_{q(4)}^B\ , \;\;\;\;\;\;\;\;\;\;\;\;\;\;\;\;\;\;\;\;\;\;\;\;\;\;\;\;\;
O_{q(4)}= O_{q(4)}^B\ , \label{o4ren} 
\end{eqnarray}
where, for simplicity, the mixing with the EOM operators as well as the BRST-exact operators is not shown, as their physical matrix elements vanish and they do not affect our final result~\cite{Kodaira:1998jn}.
Here, $O_g$, as well as $O_q$, is a mixture of the twist-two and -four operators,
and the corresponding twist-four components receive the contributions of the twist-four operators $O_{g(4)}$ and $O_{q(4)}$.
The two formulas of (\ref{o4ren}) reflect, respectively, that the twist-four operator $O_{g(4)}$ mixes with itself and another twist-four operator $O_{q(4)}$, and that $O_{q(4)}$ is renormalization group (RG)-invariant (see \cite{Tarrach:1981bi,Hatta:2018sqd,Tanaka:2018nae}).

Subtracting the traces from both sides of the equations~(\ref{o3ren}), 
$O_k$ and $O_k^B$ are, respectively, replaced by the corresponding twist-two parts, $O_{k(2)}$ and $O^B_{k(2)}$,
such that the twist-four contributions drop out. The renormalization constants $Z_T, Z_L,  Z_\psi$ and $Z_Q$
remain in the resulting equations that represent the flavor-singlet mixing of the twist-two spin-2 operators,
and thus can be determined by the second moments of the DGLAP splitting functions which are known up to the three-loop
accuracy~\cite{Vogt:2004mw}. 

For the renormalization mixing (\ref{o4ren}) at twist four,
the Feynman diagram calculation of $Z_F$ and $Z_C$ is available to the two-loop order~\cite{Tarrach:1981bi}.
Moreover, it is shown \cite{Tanaka:2018nae} that the constraints imposed by the RG invariance of  (\ref{111})
allow to determine the form of $Z_F$ as well as $Z_C$ in the MS-like schemes,
completely from $\beta(g)$ and 
$\gamma_m(g)$,  which are known to five- and four-loop order in the literature, respectively.

Therefore, six renormalization constants $Z_T, Z_L,  Z_\psi, Z_Q, Z_F$ and $Z_C$
among ten constants arising in (\ref{o3ren}),  (\ref{o4ren}) are
available to a certain accuracy in the MS-like schemes, and they 
take the form,  
\begin{equation}
Z_X=\left(\delta_{X,T}+\delta_{X,\psi}+\delta_{X,F}\right)+\frac{a_X}{\epsilon}+\frac{b_X}{\epsilon^2}+\frac{c_X}{\epsilon^3}
+\cdots
\ ,
\label{zx}
\end{equation}
in the $d=4-2\epsilon$ spacetime dimensions with $X=T,L, \psi , Q,F$, and $C$; here, $a_X, b_X, c_X, \ldots,$ are the constants given as power series in $\alpha_s$, and
$\delta_{X,X'}$ denotes the Kronecker 
symbol.
However, $Z_M$, $Z_S$, $Z_K$ and $Z_B$ still remain unknown.
It is shown \cite{Tanaka:2018nae} that these four renormalization constants can be determined to the accuracy same as the renormalization constants (\ref{zx}),
by invoking that they should also obey (\ref{zx}) with $X =M, S, K, B$, and that the RHS of the formulas of  (\ref{o3ren}) are, in total, UV-finite. Thus, all the renormalization constants in  (\ref{o3ren}),  (\ref{o4ren})  are determined
up to the three-loop accuracy, and this result allows us to derive the three-loop formulas~\cite{Tanaka:2018nae} for (\ref{ano}), by calculating the trace part of (\ref{o3ren}).

\section{Anomaly-induced mass structure of hadrons}

The QCD trace anomaly (\ref{111}) signals the generation of a nonperturbative mass scale, say, the nucleon mass $m_N$:
Taking the matrix element of  (\ref{111}) in terms of a hadron state $|h(p)\rangle$ with the 4-momentum $p^\mu$ as $p^2=m_h^2$, and using the fact that $\langle h(p)|T^{\mu\nu}|h(p)\rangle= 2p^\mu p^\nu$, 
we obtain
\begin{equation}
2m_h^2=\langle h(p)|T^\mu_\mu|h(p)\rangle
=\langle h(p)| \left(\frac{\beta(g)}{2g}F^2
+ \left(1+\gamma_m(g)\right)m\bar{\psi}\psi\ \right)|h(p)\rangle\ ,
\label{mass}
\end{equation}
so that almost all of the hadron mass $m_h$ could be attributed to the
quantum loop effects in QCD responsible for the trace anomaly.
Based on (\ref{mass}) for the nucleon ($h=N$), it is frequently argued that
the entire mass $m_N$ comes from gluons in the chiral limit.
However, the partition of QCD loop effects as
(\ref{ano}) shows
that the latter statement would not be suitable:
Indeed,
(\ref{ano}) allows us to separate (\ref{mass}) as 
\begin{equation}
2m_h^2=\langle h(p)| \eta_{\lambda\nu}T^{\lambda\nu}_g(\mu)|h(p)\rangle+ \langle h(p)| \eta_{\lambda\nu}T^{\lambda\nu}_q(\mu)|h(p)\rangle\ ,
\label{muindependentsum}
\end{equation}
and, evaluating (\ref{ano}) with $N_c=3$, $n_f=3$ at the renormalization scale $\mu$, one finds~\cite{Tanaka:2018nae}
\begin{eqnarray}
&&\eta_{\lambda\nu}T^{\lambda\nu}_g(\mu)=\left( -0.437676
   \alpha _s(\mu)
   -0.261512 \alpha _s^2(\mu)-0.183827 \alpha _s^3(\mu)\right)
   F^2(\mu) +\cdots\ ,
\label{tg3loop1}
   \\
&&\eta_{\lambda\nu}T^{\lambda\nu}_q(\mu) = \left( 0.0795775 \alpha _s(\mu)
+0.0588695 \alpha _s^2(\mu)+0.0216037 \alpha_s^3(\mu)\right)
F^2(\mu)
+\cdots\ ,
\label{tq3loop1}
\end{eqnarray}
with the ellipses associated with the operator
$m\bar \psi \psi$.
The nucleon ($h=N$) in the chiral limit gives
\begin{equation}
\frac{\langle N(p)|\eta_{\lambda\nu}T^{\lambda\nu}_q(\mu)|N(p)\rangle}{\langle N(p)| \eta_{\lambda\nu}T^{\lambda\nu}_g(\mu)|N(p)\rangle}=
-0.181818-0.0258682 \alpha _s(\mu)+ 0.0424613 \alpha _s^2(\mu)\ ,
\label{property}
\end{equation}
where $\frac{n_f}{3}/(- \frac{{11{C_A}}}{6})\cong-0.181818$.
Eqs.~(\ref{muindependentsum})-(\ref{property}), combined with $\langle N(p)|F^2|N(p)\rangle <0$,
show that the gluon- and quark-loop effects make the nucleon mass 
heavy and light, respectively, with the magnitude of the former being five times larger 
than that of the latter. From (\ref{property}), the $\mu$-dependence of this result for the relative size 
of the gluon/quark loop effects in the chiral limit is rather weak.
It is also worth noting that the total sum (\ref{muindependentsum}) of (\ref{tg3loop1}) and (\ref{tq3loop1}) 
allows us to constrain the matrix element of $F^2$ as
$\langle N(p)|F^2(\mu=1~{\rm GeV})|N(p)\rangle\simeq-8.61m_N^2$,
using
$\alpha_s(1~\rm{GeV})= 0.47358
\ldots$,
as the three-loop running coupling constant in the $\overline{\rm MS}$ scheme with
$\alpha_s(M_Z)=0.1181$. 
We note that the neglected four-loop contributions
are expected to produce corrections less than ten percent because 
$\alpha^3_s(1~\rm{GeV})\simeq0.1$.

Next, we consider the pion case, for which the PCAC relation,
$- \left(m_u + m_d\right)\langle 0|\left(\bar{u}u +\bar d d \right) |0\rangle = 2f_\pi^2m_\pi^2$,
with $f_\pi$ the pion decay constant,
indicates $m_\pi^2 \sim m$ as $m\to 0$. 
Eq.~(\ref{mass})  for the pion ($h=\pi$)
implies,
$\left. \bigl\langle \pi(p)\bigl| F^2\bigr|\pi(p)\bigr\rangle\right|_{m\to0}=0$, in the chiral limit $m\to 0$.
Eq.~(\ref{mass}) also gives the relation 
among the ${\cal O}(m)$ terms:
When the substitution, $\bigr|\pi(p)\bigr\rangle \to \bigr|\pi(p)\bigr\rangle_0+\bigr|\pi(p)\bigr\rangle_1+\ldots$,
is made,
where $\bigr|\pi(p)\bigr\rangle_0\equiv  \left. \bigr|\pi(p)\bigr\rangle\right|_{m=0}$ and $\bigr|\pi(p)\bigr\rangle_1$ 
is the ${\cal O}(m^1)$-term, 
we have,
$\bigl\langle \pi(p)\bigl|m\bar{\psi} \psi \bigr|\pi(p)\bigr\rangle 
\to\  _{0}\bigl\langle \pi(p)\bigl|  m\bar{\psi} \psi
 \bigr|\pi(p)\bigr\rangle_0$
and 
$\bigl\langle \pi(p)\bigl| F^2\bigr|\pi(p)\bigr\rangle\to\ _{0}\bigl\langle \pi(p)\bigl|  F^2\bigr|\pi(p)\bigr\rangle_1
+\ _{1}\bigl\langle \pi(p)\bigl|  F^2\bigr|\pi(p)\bigr\rangle_0$, up to the corrections of ${\cal O}(m^2)$.
The pion mass can also be calculated as the mass shift due to the ordinary
first-order perturbation theory in the quark mass term in the QCD Hamiltonian,
as~\cite{Gasser:1982ap}
\begin{equation}
m_\pi^2=\ _{0}\bigl\langle \pi(p)\bigl| m\bar{\psi} \psi
 \bigr|\pi(p)\bigr\rangle_0\ ,
\label{chiralp}
\end{equation}
and, combining this with the above results for the ${\cal O}(m)$ terms in (\ref{mass}), we obtain
\begin{equation}
\bigl(1-\gamma_m(g)\bigr)m_\pi^2
=\bigl\langle \pi(p)\bigl|  \frac{\beta (g)}{2g}F^2\bigr|\pi(p)\bigr\rangle
\ ,
\label{masspi3}
\end{equation}
to the ${\cal O}(m)$ accuracy. Therefore, up to the corrections of ${\cal O}(m^2)$, the terms associated with the $F^2$ operator and the $m\bar \psi \psi$ operator  in the RHS
of  (\ref{mass})
contribute to $m_\pi^2$ according to the relative weights,  $\left(1-\gamma_m(g)\right)$ and
$\left(1+\gamma_m(g)\right)$, respectively; here, 
$\gamma_m(g)= 0.63662 \alpha _s+ 0.768352 \alpha _s^2+ 0.801141 \alpha _s^3\simeq 0.559$,
at the three-loop accuracy.
Substituting (\ref{chiralp}) and (\ref{masspi3}) into
(\ref{muindependentsum})
with (\ref{ano}), 
we find
\begin{eqnarray}
&&\!\!\!\!\!\!\!\!\!\!\!\!\!\!\!\!\!\!\!\!\!\!\!\!
\frac{1}{2m_\pi^2}\bigl\langle \pi(p)\bigl|\eta_{\lambda\nu}T^{\lambda\nu}_g
(\mu)\bigr|\pi(p)\bigr\rangle
=0.611111-0.12215 \alpha
   _s(\mu)-0.124659 \alpha _s^2(\mu)
-0.0430357 \alpha _s^3(\mu)\ ,
\\
&&\!\!\!\!\!\!\!\!\!\!\!\!\!\!\!\!\!\!\!\!\!\!\!\!
\frac{1}{2m_\pi^2}\bigl\langle \pi(p)\bigl|
\eta_{\lambda\nu}T^{\lambda\nu}_q(\mu)\bigr|\pi(p)\bigr\rangle
=0.388889+0.12215 \alpha _s(\mu)+0.124659 \alpha _s^2(\mu)
+0.0430357 \alpha _s^3(\mu)
 \ ,
\end{eqnarray}
and, using
$\alpha_s(1~\rm{GeV})= 0.47358
\ldots$, we obtain
\begin{equation}
\frac{1}{2m_\pi^2}\bigl\langle \pi(p)\bigl|\eta_{\lambda\nu}T^{\lambda\nu}_g(\mu=1~{\rm GeV})\bigr|\pi(p)\bigr\rangle=0.521\ ,
\;\;\;
\frac{1}{2m_\pi^2}\bigl\langle \pi(p)\bigl|
\eta_{\lambda\nu}T^{\lambda\nu}_q(\mu=1~{\rm GeV})\bigr|\pi(p)\bigr\rangle=0.479\ ,
\label{eq19}
\end{equation}
which hold to the ${\cal O}(m)$ accuracy. Again, the $\mu$-dependence of the result is rather weak,
but (\ref{eq19}) shows the structure different completely from the nucleon case:
Both the gluon- and quark-loop effects 
produce roughly half of the pion mass.
This may be a particular nature as a Nambu-Goldstone
boson.

\section{Anomaly constraints on the nucleon's twist-four gravitational form factor}

The nucleon matrix element of  each term in (\ref{tqg})
is parameterized as ($|p\rangle\equiv |N(p)\rangle$)
\begin{equation}
 \langle p'|T_{q,g}^{\mu\nu}|p\rangle = \bar{u}(p')\Bigl[A_{q,g} (t)\gamma^{(\mu}\bar P^{\nu)}
 +B_{q,g}(t)\frac{\bar P^{(\mu}i\sigma^{\nu)\alpha}\Delta_\alpha}{2m_N}
 + D_{q,g}(t)\frac{\Delta^\mu\Delta^\nu -\eta^{\mu\nu}t}{4m_N} + \bar{C}_{q,g}(t)m_N\eta^{\mu\nu}\Bigr] u(p)\ ,
\label{para}
\end{equation}
in terms of the gravitational form factors $A_{q,g} (t), B_{q,g}(t), D_{q,g}(t)$, and $\bar{C}_{q,g}(t)$~\cite{Polyakov:2018zvc,Tanaka:2018wea},
where $\Delta=p'-p$, $\bar{P}=\left(p+p'\right)/2$, 
$t=\Delta^2$, $\bar{P}^2= m_N^2 -t/4$, and $u(p)$ is the nucleon spinor.
$A_{q,g} (t)$ and $B_{q,g}(t)$ are familiar twist-two form factor,
obeying the forward ($t \to 0$) sum rules,  
$A_q(0)+A_g(0)=1$, $[A_q(0)+B_q(0) +A_g(0) +B_g(0)]/2=1/2$, 
representing a sharing of the total momentum and total angular momentum by the quarks/gluons.
$D_{q,g}(t)$, $\bar{C}_{q,g}(t)$ have also received considerable attention recently~\cite{Burkert:2018bqq,Polyakov:2018zvc,Tanaka:2018wea,Kumano:2017lhr}:
$D_{q,g}(t)$ are related to 
the so-called D term, $D\equiv D_q(0)+D_g(0)$~\cite{Polyakov:2018zvc}.
For $\bar{C}_{q,g}(t)$, 
exact manipulations for the divergence of  (\ref{tqg}) yield the operator identities,
$\partial _\nu T_q^{\mu \nu } =  \bar \psi g{F^{\mu \nu }}{\gamma _\nu }\psi$, 
$\partial _\nu T_g^{\mu \nu } =  - F_a^{\mu \nu }D_{ab}^\rho F_{\rho \nu }^b$,
up to the terms that vanish by 
the EOM, 
so that
$\bar{C}_{q}(t)=-\bar{C}_{g}(t)$, $\bar{C}_q(t)\propto \langle p'|\bar{\psi}ig F^{\mu \nu}\gamma_\nu \psi|p \rangle$,
and $\bar{C}_g(t) \propto \langle p'|F_a^{\mu \nu} iD^\rho_{ab}  F^b_{\rho \nu}|p\rangle$
hold~\cite{Tanaka:2018wea}, showing that $\bar{C}_{q,g}(t)$ are of twist four.
We note that $\bar{C}_{q,g}(t)$ are relevant to
the force distribution inside the nucleon~\cite{Polyakov:2018zvc,Hatta:2018ina}
and the nucleon's 
transverse spin sum rule~\cite{Hatta:2012jm}.

To see the consequence of  (\ref{ano}) on $\bar{C}_{q,g}(t)$,
we take the trace of the forward limit, $\Delta^\mu \to 0$, of (\ref{para}):
\begin{equation}
{\bar C}_q(t=0)=-{\bar C}_g(t=0)= -\frac{1}{4}A_q(t=0) +\frac{1}{8m_N^2}\langle p|\eta_{\lambda \nu}T_q^{\lambda \nu}|p\rangle\ ,
\label{trace}
\end{equation}
and substitute (\ref{ano}) into the second term, and the three-loop DGLAP evolution
into $A_q(\mu)\equiv A_q\left(t=0, \mu\right)$ as a flavor-singlet spin-2 operator renormalized at the scale $\mu$.
$\bar{C}_q(\mu) \equiv \bar{C}_q(t=0, \mu))$
of (\ref{trace}) 
reads~\cite{Hatta:2018sqd}
\begin{equation}
\bar{C}_q(\mu)=
- \frac{1}{4} \left( \frac{n_f}{4C_F+n_f} + \frac{2n_f}{3\beta_0}\right) + \left(\frac{2 n_f}{3 \beta _0}+1\right)\frac{\left\langle p \right| m \bar{\psi }\psi  \left| p \right\rangle}{8m_N^2}
   -\frac{4C_F A_{q0}+n_f \left(A_{q0}-1\right)}{4\left(4 C_F+n_f\right)}\left(\frac{\alpha _s\left(\mu
   \right)}{\alpha _s(\mu_0 )}\right)^{\frac{ 8 C_F+2n_f}{3 \beta_0}}+\cdots\ ,
   \label{asy}
\end{equation}
with $\beta_0\equiv(11C_A-2n_f)/3$, where $\mu_0$ is a certain starting scale, $A_{q0}\equiv A_q( \mu_0 )$, and
$\left\langle p \right| F^2   \left| p \right\rangle$ has been eliminated in favor of the nucleon mass $m_N$
using (\ref{mass}).
Here, explicitly shown are the leading order (LO) terms that are derived from the
contributions at the one-loop accuracy in (\ref{trace}); the ellipses denote the NLO and NNLO terms derived
from the two- and three-loop contributions in (\ref{trace}), respectively.
The first few terms independent of $\mu$ in (\ref{asy}) represents the asymptotic value which is completely determined by the values of  $N_c$ and $n_f$ 
in the chiral limit. Substituting $N_c=3$ and $n_f=3$, we obtain
\begin{eqnarray}
&&\!\!\!\!\!\!\!\!\!\!\!\!\!\!\!\!\!\!\!\!\!\!\!
{\left. {\bar C_q(\mu )} \right|_{{n_f} = 3}} =  \left.- 0.146 + 0.306\frac{{\langle p|m\bar qq\left| p \right\rangle }}{{2{m_N^2}}} + \left(0.09-0.25 A_{q0} \right){\left( {\frac{{{\alpha _s}\left( \mu  \right)}}{{{\alpha _s}({\mu _0})}}} \right)^{\frac{{50}}{{81}}}}
 + {\alpha _s}(\mu )\ \right( 0.006 
 \nonumber\\
 && \!\!\!\!\!\!\!\!\!\!\!\!\!\!\!\!\!\!\!\!\!\!\!
 + \left. 0.08\frac{{\langle p|m\bar qq\left| p \right\rangle }}{{2{m_N^2}}}{\rm{ }} +\left( 0.013- 0.035A_{q0} \right){{\left( {\frac{{{\alpha _s}\left( \mu  \right)}}{{{\alpha _s}({\mu _0})}}} \right)}^{\frac{{50}}{{81}}}} 
 +\left( 0.035A_{q0} -0.028 \right){{\left( {\frac{{{\alpha _s}\left( \mu  \right)}}{{{\alpha _s}({\mu _0})}}} \right)}^{-\frac{{31}}{{81}}}}   \right) + \cdots\ ,
\label{ccc}
\end{eqnarray}
where the NLO as well as LO terms are explicitly shown, and the ellipses stand for the NNLO terms. 
For illustration, we plot (\ref{ccc}) as a function of $\mu$ in the chiral limit: Fig.~1(a) shows the results up to the LO, NLO, and NNLO accuracy;
the NLO as well as NNLO corrections give a
few percent level effects, reflecting the small numerical coefficients arising in (\ref{ccc}),
and
the NLO and NNLO corrections tend to cancel; the apprach to the asymptotic value, $-0.146$, is quite slow. The NNLO result is separated, in Fig.~1(b), into the individual contributions of each term in (\ref{trace}), the first (twist-2) term and the second (anomaly) term; both  twist-2 and anomaly terms produce the important effects.

To summarize, the quark/gluon 
decomposition of the QCD trace anomaly allows us to study the hadrons from new aspects, revealing e.g., quite different pattern
between the nucleon and the pion.

\vspace{-0.5cm}
\begin{figure}[hbtp]
\begin{center}
\includegraphics[width=0.4\textwidth]{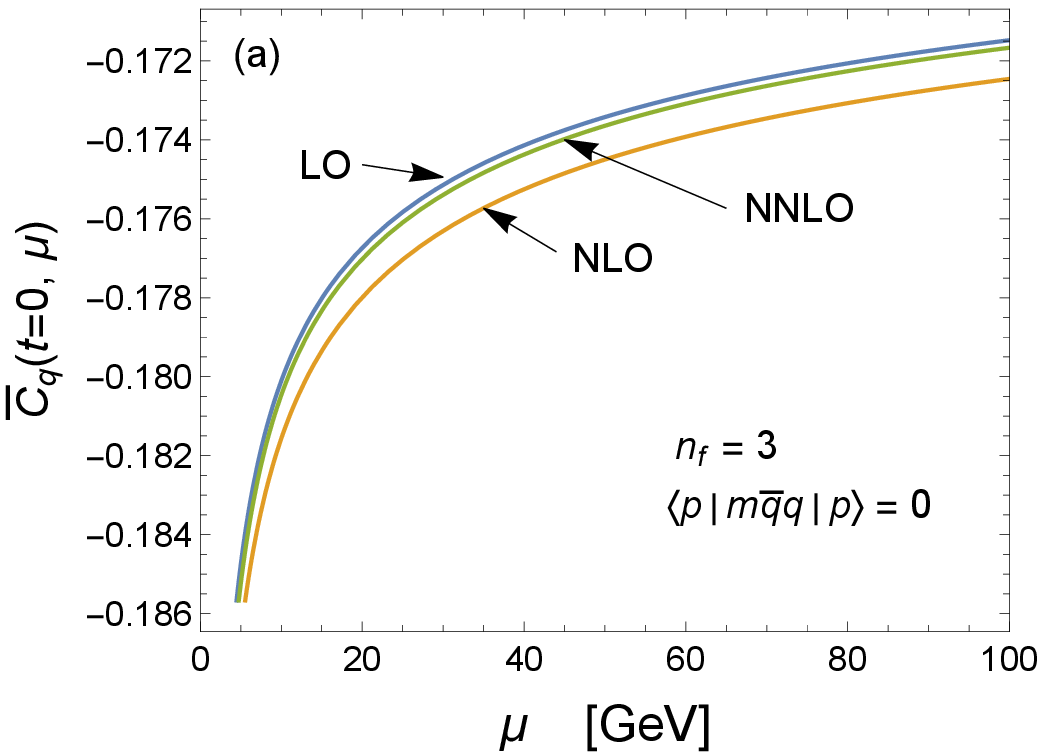}
\hspace{1.2cm}
\includegraphics[width=0.4\textwidth]{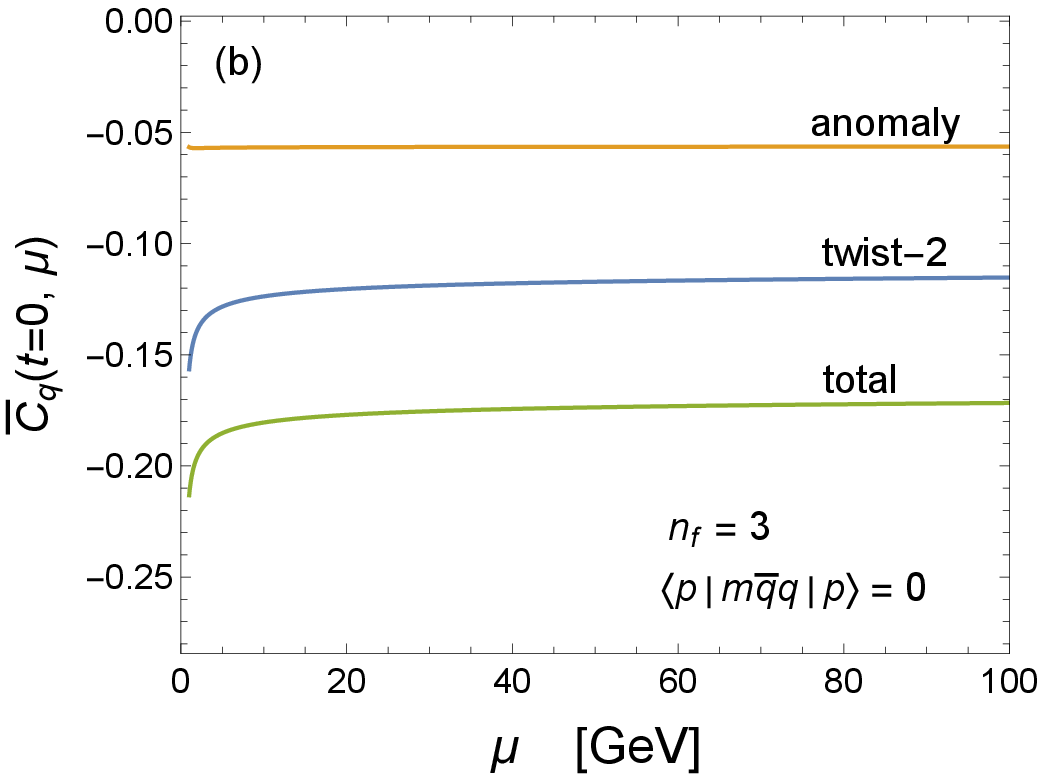}
\caption{The 
nucleon's gravitational form factor ${\bar C}_q(t=0, \mu )$ at the NNLO (3-loop) accuracy:
(a) the LO, NLO, and NNLO calculations; (b) the contributions of the first (twist-2) and second (anomaly) terms of (\ref{trace}).
}
\label{fig1}
\end{center}
\end{figure}

\vspace{-0.7cm}
\noindent
{\bf Acknowledgments:}
This work was supported by JSPS KAKENHI Grant Number JP19K03830.


\begin{thebibliography}{9}


\bibitem{Collins:1976yq}
J.~C.~Collins, A.~Duncan and S.~D.~Joglekar,
Phys. Rev. D \textbf{16}, 438-449 (1977).

\bibitem{Hatta:2018sqd} 
  Y.~Hatta, A.~Rajan and K.~Tanaka,
  JHEP {\bf 1812}, 008 (2018).
  
\bibitem{Tanaka:2018nae} 
  K.~Tanaka,
  JHEP {\bf 1901}, 120 (2019).

\bibitem{Kodaira:1998jn}
J.~Kodaira and K.~Tanaka,
Prog. Theor. Phys. \textbf{101}, 191-242 (1999).



\bibitem{Tarrach:1981bi}
R.~Tarrach,
Nucl. Phys. B \textbf{196}, 45-61 (1982).


\bibitem{Vogt:2004mw}
A.~Vogt, S.~Moch and J.~A.~M.~Vermaseren,
Nucl. Phys. B \textbf{691}, 129-181 (2004).



\bibitem{Gasser:1982ap}
J.~Gasser and H.~Leutwyler,
Phys. Rept. \textbf{87}, 77-169 (1982).


\bibitem{Burkert:2018bqq} 
  V.~D.~Burkert, L.~Elouadrhiri and F.~X.~Girod,
  Nature {\bf 557}, no. 7705, 396 (2018).

\bibitem{Polyakov:2018zvc} 
  M.~V.~Polyakov and P.~Schweitzer,
  Int.\ J.\ Mod.\ Phys.\ A {\bf 33}, no. 26, 1830025 (2018).

\bibitem{Tanaka:2018wea} 
  K.~Tanaka,
  Phys.\ Rev.\ D {\bf 98}, no. 3, 034009 (2018).



\bibitem{Kumano:2017lhr} 
  S.~Kumano, Q.~T.~Song and O.~V.~Teryaev,
  Phys.\ Rev.\ D {\bf 97}, no. 1, 014020 (2018).

  
\bibitem{Hatta:2018ina} 
  Y.~Hatta and D.~L.~Yang,
  Phys.\ Rev.\ D {\bf 98}, no. 7, 074003 (2018).
  
  
 \bibitem{Hatta:2012jm} 
  Y.~Hatta, K.~Tanaka and S.~Yoshida,
  JHEP {\bf 1302}, 003 (2013).
  


\end{thebibliography}
\end{document}